\documentclass[conference]{IEEEtran}
\IEEEoverridecommandlockouts
\usepackage{cite}
\usepackage{amsmath,amssymb,amsfonts}
\usepackage{algorithmic}
\usepackage{graphicx}
\usepackage{textcomp}
\usepackage{xcolor}
\usepackage{comment}
\usepackage{multirow, array}
\usepackage{marvosym}
\usepackage{tabularx}
\usepackage{booktabs}
\usepackage{longtable}
\usepackage{caption}
\usepackage[hyphens,spaces,obeyspaces]{url}
\usepackage{hyperref}
\usepackage{balance}
\def\BibTeX{{\rm B\kern-.05em{\sc i\kern-.025em b}\kern-.08em
    T\kern-.1667em\lower.7ex\hbox{E}\kern-.125emX}}

\begin{document}

\title{Wicked Problem, Parsimonious Solution: Securing Electric Vehicle Charging Station Software}

\author{\IEEEauthorblockN{Emma Sheppard\IEEEauthorrefmark{1}\IEEEauthorrefmark{2},
Zachary Wadhams\IEEEauthorrefmark{1},
Dalton Arford\IEEEauthorrefmark{2},
Clemente Izurieta\IEEEauthorrefmark{1}\IEEEauthorrefmark{2}\IEEEauthorrefmark{3},
and
Ann Marie Reinhold\IEEEauthorrefmark{1}\IEEEauthorrefmark{2}\\[3.0ex]
\IEEEauthorblockA{\IEEEauthorrefmark{1}Montana State University, Bozeman MT, USA}
\IEEEauthorblockA{\IEEEauthorrefmark{2}Pacific Northwest National Laboratory, Richland WA, USA}
\IEEEauthorblockA{\IEEEauthorrefmark{3}Idaho National Laboratory, Idaho Falls ID, USA}}}

\maketitle

\begingroup
\renewcommand\thefootnote{}
\footnotetext{%
\footnotesize
\copyright{} 2025 IEEE. Personal use of this material is permitted.
Permission from IEEE must be obtained for all other uses, in any current or future media,
including reprinting/republishing this material for advertising or promotional purposes,
creating new collective works, for resale or redistribution to servers or lists,
or reuse of any copyrighted component of this work in other works.
\par\medskip
This is the author's accepted version of:
E. Sheppard, Z. Wadhams, D. Arford, C. Izurieta, and A. M. Reinhold,
``Wicked Problem, Parsimonious Solution: Securing Electric Vehicle Charging Station Software,''
2025 IEEE International Conference on Cyber Security and Resilience (CSR),
Chania, Greece, 2025, pp. 679--686.
doi: \url{https://doi.org/10.1109/CSR64739.2025.11130101}.
}
\endgroup

\begin{abstract}

Electric vehicle charging infrastructure presents a suite of novel cyber-physical threats.
Among this infrastructure, charging stations are the most vulnerable elements.
The software in the charging station supply equipment is particularly vulnerable.
Currently, the software is an attack surface that is largely unprotected and poorly characterized.
To represent the vulnerabilities in this attack surface, we advocate for applying modern software quality assurance to characterize vulnerabilities in electric vehicle charging station software.
Specifically, we advocate for the application of hierarchical software quality assurance (HSQA) to specialized electric vehicle charging station software.
HSQA provides a comprehensive view of the code quality and security --- from the level of individual vulnerabilities (e.g., CVEs) to high level characteristics (e.g., CIA Triad).
HSQA incorporates quality and security considerations throughout the software development lifecycle. 
Thus, our position is that HSQA is an excellent approach for assessing electrical vehicle charging station software.

\end{abstract}


\section{Introduction}

Worldwide, there is a growing demand for carbon-neutral technology, with many nations moving towards the adoption of zero-emissions vehicles \cite{b1, b2, b3, b4}. Consequently, electric vehicle charging infrastructure (EVCI) is rapidly growing. EVCI refers to every component required to charge an electric vehicle (EV), including the charging station (EVCS), the power supply equipment (EVSE) within the charging station, the cloud-based charging station management systems (CSMS), the power grid or operator, and the payment and authorization mechanisms \cite{b5}. 

EVCI creates a nexus between the EV, the cloud, and the power grid. Thus, EVCI is a complex system that qualifies as both an Industrial Internet of Things (IIOT) device and operational technology (OT), and incorporates components of industrial automation and control systems (IACS), as well as distributed energy resources (DER) \cite{b5}. 

Public EVCI (i.e., commercial charge points) supports the growing EV market by providing accessibility to both rural and urban areas, alleviating range anxiety and encouraging the adoption of EVs \cite{b4,b9}. The push for the public adoption of EVs will require substantial investments in public EVCI to support the growing EV market \cite{ormazabal_ev_investment, b10}. This expansion of public EVCI will introduce myriad logistical challenges, such as access and reliability, and security concerns, including cybersecurity threats. 

\section{The Attack Surface of EVCI}

As the demand for EVs has increased---and continues to increase---around the globe, manufacturers are increasing public access to EVCI. With increasing access comes an expanding attack surface\footnote{https://www.fortinet.com/resources/cyberglossary/attack-surface}. This attack surface is concerning to cybersecurity researchers and original equipment manufacturers (OEMs) alike, as EVCS equipment is public facing by necessity. The attack surface of EVCI is expansive and highly varied across charging stations---depending upon the combinations of proprietary and open-source software, hardware, and protocols within the infrastructure \cite{b11}. Existing attempts to characterize this attack surface are limited. No one study encompasses every component of every EVCS due to the variation in hardware and software in use.

An EVCS contains one or more EVSEs. The EVSE is the most vulnerable element of the infrastructure because it is public facing and central to the transmission of data and power. The EVSE is the hub for the exchange of information between the EV, the EV user, the third-party application providers, the OEM, the charge point operator (CPO), and the grid operator \cite{b13}. Thus, threat modeling of the attack surface in EVCI should include a comprehensive assessment of the EVSE \cite{b13, b14}. 

EVSEs are sophisticated cyber-physical systems (CPS) that leverage interconnected technologies, enabling them to manage the flow of electricity from the grid to the vehicle and facilitate data exchange between users and monitoring or management systems. Given its crucial role in the core functionality of EVCI, securing the EVSE is paramount to ensuring the security of the entire infrastructure.

The attack surface of EVSEs can be broken into two categories: a physical surface, composed of hardware, and a digital surface, driven by software. With the automotive industry moving towards “software-defined vehicles,” protecting the digital attack surface is crucial \cite{b48}. Software governs key interdependent functionalities of the EVSE, making it a prime target for adversarial attacks \cite{b49}. By focusing on the software that drives the EVSE, attacks can be mitigated across its interfaces. Identifying and improving software quality and security characteristics are critical for reducing the exploitability of EVCI. 

\textit{Here, we explore a novel integration and assessment of EVSE software quality and security characteristics using a tried-and-true software quality assurance approach.}


\subsection{Prior Work}
\subsubsection{Four-Interface Threat Model}

The attack surface of the EVSE can be categorized into four interfaces: EV-to-EVSE, EV operator, EVSE internet, and EVSE maintenance \cite{b13}. Each of these four interfaces are explored below. 

Interface 1: The EV-to-EVSE interface is the EVSE coupling cable. This cable physically connects the EV to the EVSE for power exchange. This interface is not standardized. EVSE couplers vary in power level, power type, and underlying communication protocols, resulting in a vulnerable heterogeneous infrastructure. Vulnerabilities in this interface include malware exchange from EV to EVSE \cite{b15}, charging disruptions \cite{b16,b17}, privilege escalation within Vehicle-to-Grid communications \cite{b18}, and security concerns with the ISO 15118 EV-to-EVSE communication protocol \cite{b19,b20,b21,b22}. 

Interface 2: The EV operator interface authenticates charging sessions using methods like Radio Frequency Identification (RFID) tags, smartphone Near Field Communication (NFC), and credit card swipes to link the operator’s billing information to the charging station. Vulnerabilities in the EV operator interface include RFID cloning \cite{b23,b24}, authorization bypass mechanisms \cite{b24}, and reverse-engineering of third-party applications to gain access to EVSE management portals \cite{b25}. Note that ISO 15118-20\footnote{https://www.iso.org/standard/77845.html} recommends public key infrastructure (PKI) encryption to mitigate this attack vector\cite{b26}. 

Interface 3: The EVSE internet interface encompasses any vulnerability that may occur due to EVSE connection to internet services. EVSEs are required to maintain internet connectivity to transmit telemetry data, allow access for third-party management systems, and enable grid operators to access EVSE equipment \cite{b13}. EVSE communication protocols tend to be proprietary, but open-source protocols exist. 

Open Charge Point Protocol (OCPP) is a common open-source protocol used for communication between the EVSE and its CSMS \cite{b27}. A benefit of current OCPP versions (2.1 and 2.0.1) is the inclusion of PKI encryption and ISO 15118 plug-and-charge functionality. However, OCPP 1.6 is more commonly used in public charging stations \cite{b28}. This older version of OCPP lacks PKI encryption and requires the use of Virtual Private Networks (VPNs) to protect the EVSE from man-in-the-middle (MITM) attacks and energy theft \cite{b29}. 

Attacks on this interface can target EVSE vendors, grid operator systems, and the EV \cite{b13}. Documented exploits include targeting OCPP attack vectors \cite{b24,b29,b31}, intercepting billing communications \cite{b23}, and detecting EVSEs on the public internet \cite{b32,b33}. Remote communications with an EVSE can allow an attacker to gain remote access to other EVSEs \cite{b34}. 

Interface 4: The EVSE maintenance interface is the physical, outward-facing hardware. Communications in the maintenance interface circuit boards generally occur over ethernet or serial analog and are often unencrypted \cite{b35}. Ethernet switches can be accessed if the lid to the EVSE is removed, and communications amongst components can be monitored using an ethernet cable. Outside the lid, physical ports are often left unprotected to allow vendors to monitor, update, and debug components within the EVSE, resulting in an easily exploitable attack vector. Moreover, the maintenance interface includes locally hosted web servers through which adversaries can gain access to personally identifiable information (PII) \cite{b36}. Additional hazards include unsigned firmware \cite{b25} and hard-coded credentials  \cite{b38,b39}.

\subsubsection{CharIN EVSE Threat Model}

Charging Interface Initiative e.V. (CharIN) released the most current and comprehensive characterization of the EVSE attack surface \cite{b14}, enumerating several attack vectors and threat scenarios. The CharIN EVSE model includes a mapping of each scenario to the Microsoft STRIDE threat model\footnote{https://learn.microsoft.com/en-us/azure/security/develop/threat-modeling-tool-threats} and recommends mitigations and best practices. Referenced frameworks, standards, and protocols include ISA/IEC 62443 Cybersecurity for ICAS\footnote{https://www.isa.org/standards-and-publications/isa-standards/isa-iec-62443-series-of-standards}, MITRE ATT\&CK Framework\footnote{https://attack.mitre.org/matrices/ics/}, ISO 15118-2\footnote{https://www.iso.org/standard/55366.html} and ISO 15118-20\footnote{https://www.iso.org/standard/77845.html}, OCPP\footnote{https://openchargealliance.org/protocols/open-charge-point-protocol/}, and ISO/SAE 21434: Road Vehicles – Cybersecurity Engineering\footnote{https://www.sae.org/standards/content/iso/sae21434/}.

\subsubsection{Integrating Known Threat Models for Comprehensive EVCI Security}

The Four-Interface \cite{b13} and CharIN EVSE threat models \cite{b14} highlight the criticality of both software and hardware components within the EVSE. Both threat models are applicable to EVSE components across OEMs. Here, we take a position that incorporates findings from both models, addresses exploitable attack vectors, and is adaptive to emerging threats and vendor-dependent components. 

\section{Our Position on Securing EVSE Software}

\begin{figure*}
    \centering
    \includegraphics[width=1\linewidth]{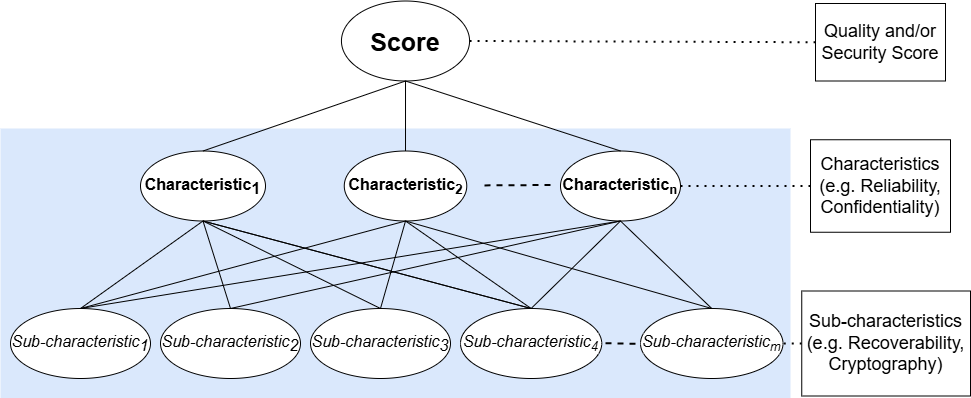}
    \caption{A conceptual view of the upper-level architecture of an HSQA model. Includes the overall quality and/or security score, high-level characteristics, and their sub-characteristics. For more examples, see Tables I, II, and III.}
    \label{fig:enter-label}
\end{figure*}
\raggedbottom

Utilizing secure-by-design principles \cite{b50}, we advocate for an approach that incorporates security considerations throughout the software development lifecycle. Historically, charging stations and their software have been constructed with a ``build-and-forget” mentality \cite{b14}. These practices require practitioners to shoulder the burden of maintaining externally developed software \cite{b50}. 

More recently, cybersecurity agencies are urging manufacturers to assume full responsibility for their software \cite{b50}. Technical experts are moving toward standardizing cybersecurity practices throughout an EVSE rather than inheriting security measures from individual vendor components \cite{b14}. For instance, CharIN is pushing for EVSE software components to be secure-by-design \cite{b14}. This change from ``build-and-forget" to secure-by-design is a shift from a reactive to a proactive security posture. 

A proactive security posture requires protecting EVSE software. Protecting EVSE software can be achieved through quality and security evaluation methods, such as software quality assurance (SQA). SQA models---and specifically, hierarchical SQA models (HSQA)---enable the identification, prioritization, and mitigation of security issues \cite{b51}. HSQA modeling answers the calls by the U.S. Department of Homeland Security Cybersecurity \& Infrastructure Security Agency (DHS CISA) and CharIN for incorporation of secure-by-design principles \cite{b52, b53}. Our position is that HSQA is a promising avenue to evaluate the quality and security of EVSE software. In the following subsections, we advocate for rigorous assessment of software quality and security using HSQA to promote a more secure and resilient EVSE.

\subsection{Quality Modeling for Software Components in the EVSE}

We propose HSQA to promote the quality and security of EVSE software. We incorporate existing work in IIOT, IACS, DER, OT, and the automotive industry (e.g., ISO/IEC 33000\footnote{https://committee.iso.org/sites/jtc1sc7/home/projects/flagship-standards/isoiec-33000-family.html} and SPICE \cite{b56}) into HSQA. HSQA allows for the generalization of quality in EVSE software. 

HSQA has been employed successfully for over a decade \cite{b58, b59}. Initially, these models operationalized ISO/IEC 9126:2001\footnote{https://www.iso.org/standard/22749.html}, the precursor to ISO/IEC 25010:2011 and ISO/IEC 25010:2023\footnote{https://www.iso.org/standard/78176.html}. The modern approach to HSQA is the Platform for Investigative software Quality Understanding and Evaluation (PIQUE) \cite{b53}. PIQUE is domain-agnostic, allowing efficient application across various software-dependent systems, including EVSE. 

HSQA evaluates and scores software utilizing outputs from static analysis tools. These outputs are aggregated into increasingly abstract concepts; for example, from raw counts of CVEs\footnote{https://cve.mitre.org/} to high level characteristics in ISO/IEC standards \cite{reinhold2024agg, b53}. HSQA enables stakeholders to make informed decisions regarding software quality and security, supporting risk analysis by organizing potential software-level vulnerabilities across multiple levels of abstraction. 

HSQA models are customizable by design. This customizability is an advantage because EVSE software is necessarily diverse. Diverse software is analyzed by diverse static analysis tools; e.g., C\# code must be analyzed with different static analysis tools than C++ code. Yet, the information garnered from diverse tools can be integrated into HSQA models with minimal effort.

The diverse software within EVSE operate in concert and thus should not be considered in isolation. HSQA enables a single solution that can evaluate and score each software component contemporaneously. In so doing, HSQA provides a means for assessing the quality and security posture of the EVSE as a system.


Integrated assessments of software quality and security require identifying high-level quality and security characteristics specific to an EVSE. Integration is achieved by merging software quality standards (e.g., ISO/IEC 25010:2023) with EVSE cybersecurity research, standards, and best practices (e.g.,\cite{b5, b40, b43, b47, b62, b63, b64}). HSQA facilitates this merging by design (Fig. 1). 

Our position is: \textit{\textbf{HSQA is an excellent approach for the assessment of EVSE software.} This novel application of a tried-and-true solution will increase the quality and security posture of EVSE.} 

\subsection{Counterarguments}

Counterarguments to this position may include alternate ways to satisfy secure-by-design principles or evaluate the quality of EVSE software. Quality assurance and secure-by-design best practices include the use of memory-safe programming languages, vulnerability disclosures, static and dynamic application security testing (SAST/DAST), and code review for quality assurance \cite{b50}. Any implementation of these best practices aids in securing software. However, these approaches are independent and therefore do not provide a holistic perspective on software quality and security.

In contrast, HSQA leverages existing vulnerability disclosures and static analysis tools to score the quality and security of software holistically. HSQA modeling can implement published CVEs, CWEs\footnote{https://cwe.mitre.org/}, code review, and SAST/DAST. Currently, outputs from static analysis tools are the inputs to HSQA models \cite{b53}. HSQA offers a practical and scalable approach for integrating such information at multiple levels of abstraction--meeting the needs of developers and the C-Suite alike \cite{b53}. 

\section{Quality \& Security Characteristics for EVSE}

The novel application of HSQA to EVSE requires the integration of high-level quality and security characteristics from diverse sources. We propose the integration of three foundational pillars: (\textit{I}) the ISO/IEC 25010:2023 software product quality model \cite{b61}; (\textit{II}) the \textit{Government Fleet and Public Sector Electric Vehicle Supply Equipment (EVSE) Cybersecurity Best Practices and Procurement Language Report} (herein K. Harnett et al.) prepared by the U.S. Department of Transportation (DoT) Volpe Center \cite{b62}; and (\textit{III}) cybersecurity standards relevant to power electronics delivery equipment, industrial control systems (ICS), and road vehicles, including ISA/IEC 62443 \cite{b43}, IEEE 1547-3 \cite{b65}, and ISO/SAE 21434:2021 \cite{b47}. Furthermore, we draw on the methods outlined in Karnouskos et al. \cite{b55} to define the ISO/IEC 25010:2023 software quality characteristics and sub-characteristics in the context of ICS and EVSE. The following proposed high-level quality and security characteristics would be implemented in a similar architecture to Fig 1.

\subsection{Pillar I: Software Quality Standards}

\begin{table*}[]
\centering
\caption{ISO/IEC 25010:2023 Systems and software engineering — Systems and software Quality Requirements and Evaluation (SQuaRE) — Product quality model characteristics (bolded), sub-characteristics (italicized), and their proposed applications to EVSE software.}
\renewcommand{\arraystretch}{1.2} 
\setlength{\tabcolsep}{6pt} 
\begin{tabular}{|p{7cm}|p{10cm}|}
\hline
\textbf{Characteristic \& \textit{Sub-characteristics}} & \textbf{Application to EVSE Software} \\ \hline

\textbf{Functional Suitability} \newline 
\textit{Functional completeness, correctness, appropriateness} &
Ensures charging stations perform as expected to meet EV user needs, enhancing satisfaction and confidence in charging infrastructure. \\ \hline

\textbf{Performance Efficiency} \newline 
\textit{Time behavior, resource utilization, capacity} &
Impacts the speed of vehicle charging and energy efficiency, directly affecting EV users, management system operators, and power grid load. \\ \hline

\textbf{Compatibility} \newline 
\textit{Co-existence, interoperability} &
Enables communication between charging stations, EVs, the grid, and third-party applications via the cloud. \\ \hline

\textbf{Interaction Capability} \newline 
\textit{Appropriateness recognizability, learnability, operability, user error protection, user engagement, inclusivity, user assistance, self-descriptiveness} &
Aids users in navigating the charging process, including use of the human-machine interface (HMI), the authentication process, and the physical interaction of charging a vehicle. \\ \hline

\textbf{Reliability} \newline 
\textit{Faultlessness, availability, fault tolerance, recoverability} &
Minimizes downtime and enables fast recovery after service disruptions, ensuring continuous charging operations. \\ \hline

\textbf{Security} \newline 
\textit{Confidentiality, integrity, non-repudiation, accountability, authenticity, resistance} &
Protects user data and EVSE networks, preventing malicious attacks that may compromise charging infrastructure \cite{b14,b62}. \\ \hline

\textbf{Maintainability} \newline 
\textit{Modularity, reusability, analysability, modifiability, testability} &
Facilitates efficient software updates and improvements, accommodating new technologies and security measures. \\ \hline

\textbf{Flexibility} \newline 
\textit{Adaptability, scalability, installability, replaceability} &
Allows charging infrastructure to evolve with technological and regulatory changes, enabling compliance with new standards, i.e. NACS SAE J3400 \cite{b69,b70}. \\ \hline

\textbf{Safety} \newline 
\textit{Operational constraints, risk identification, fail safe, hazard warning, safe integration} &
Protects users and infrastructure from hazards like cable melting, fires, and electrical shocks, enhancing public confidence in EV adoption \cite{b40}. \\ \hline
\end{tabular}
\label{tab:evci_quality_model}
\end{table*}
\raggedbottom

The ISO/IEC 25000 suite of standards provides a comprehensive framework for evaluating software product quality, with ISO/IEC 25010 – System and software quality models serving as a key component of the ISO/IEC 2501n standards for Quality Model Division\footnote{https://iso25000.com/index.php/en/iso-25000-standards/51-iso-iec-2501n}. ISO/IEC 25010:2023 specifies nine high-level characteristics that are essential for assessing software quality: functional suitability, performance efficiency, compatibility, interaction capability, reliability, security, maintainability, flexibility, and safety \cite{b61}. Table I outlines these high-level characteristics, their sub-characteristics, and their proposed applications to EVSE software. Definitions exactly as written in the ISO/IEC 25010:2023 Systems and software engineering (SQuaRE) Product quality model can be found in the full archived table\footnote{https://doi.org/10.5281/zenodo.14758339}.

\subsection{Pillar II: Cybersecurity Best Practices for EVSE}

K. Harnett et al. and the U.S. Naval Facilities Engineering Systems Command prepared the most current EVSE best practices for cybersecurity \cite{b62}. These best practices are based on interviews with subject matter experts, the ElaadNL-commissioned European Network for Cyber Security (ENCS) \textit{EV Charging System Security Requirements} \cite{b71}, and the U.S. National Motor Freight Traffic Association cybersecurity reports for medium and heavy-duty EVs \cite{b72}. The best practices align with the Microsoft STRIDE Threat Model, which categorizes common cyber threats and maps them to security characteristics \cite{b42, stride}. 

\begin{table*}[]
\centering
\caption{Security sub-characteristics from K. Harnett et al. \cite{b62}, the corresponding requirement from the ElaadNL-commissioned ENCS \textit{EV Charging Systems Security Requirements} \cite{b71}, and their descriptions.}
\renewcommand{\arraystretch}{1.1} 
\setlength{\tabcolsep}{5pt} 
\begin{tabular}{|p{3.5cm}|p{4cm}|p{8.5cm}|}
\hline
\textbf{K. Harnett et al. Security Sub-characteristic} & \textbf{ENCS Requirement} & \textbf{ENCS Description \small $^{\text{\Cross}}$} \\ \hline

Design & Future Proof Design & Prevents lack of capabilities for future security updates. \\ \hline

Cryptography & Cryptographic Algorithms and Protocols & Describes cryptographic algorithms, key lengths, and pseudo-random generators allowed for use. \\ \hline

Communication & Communication Security & Defines implementation mechanisms for end-to-end security in an EVCS. \\ \hline

Hardening & System Hardening & Provides hardening mechanisms for the EVCS components. \\ \hline

Resiliency & Resilience & Prevents issues due to misuse of the EVCS components or communication interfaces. \\ \hline

Secure Operation & Access Control & Defines authorization mechanisms for the EVCS components or its communication interfaces. \\ \hline

Logging & Logging & Defines detection mechanisms to identify security issues on an EVCS component or its communication interfaces. \\ \hline

Assurance & Assurance & Specifies measures vendors must take to ensure secure functioning of EVCS components. \\ \hline

Lifecycle and Governance & Product Lifecycle and Governance & Defines processes for secure development, manufacturing, and provisioning of EVCS components. \\ \hline

EVSE Operator\slash Utility Operator Communications & CPO and DSO Communication & Requirements for secure communications between Charge Point Operators (CPOs) and Distribution System Operators (DSOs). Useful for new server procurement or setup. \\ \hline

\multicolumn{3}{l}{\small $^{\text{\Cross}}$ Descriptions have been paraphrased and condensed for readability. Exact descriptions can be found in Section 1.4 of \cite{b71}.}
\end{tabular}
\label{tab:encs_security}
\end{table*}
\raggedbottom

All of the STRIDE properties are important for evaluating EVSE software (i.e., authenticity, integrity, non-repudiation, confidentiality, availability, and authorization). We use the STRIDE properties as our characteristics. The following sub-characteristics are also important (Table \ref{tab:encs_security}): design, cryptography, communication, hardening, resiliency, secure operation, logging, assurance, lifecycle and governance, and EVSE operator/utility operator communications. 

\subsection{Pillar III: Cybersecurity Standards for Road Vehicles, Industrial Automation and Control Systems, and Power Delivery Electronics}

\begin{table*}[]
\centering
\caption{Security frameworks (italicized), high-level characteristics (bolded), sub-characteristics (italicized), and their definitions from ISO/SAE 21434:2021, ISA/IEC 62443, and IEEE 1547-3.}
\begin{tabular}{|l|l|l|}
\hline
\textbf{\begin{tabular}[c]{@{}l@{}}Standard \& \textit{Security Framework}\end{tabular}} & 
\textbf{\begin{tabular}[c]{@{}l@{}} Security Characteristic/\\ \textit{Sub-characteristic}\end{tabular}} & 
\textbf{Definition} \\ \hline

\begin{tabular}[c]{@{}l@{}}\textbf{ISO/SAE 21434:2021 Road Vehicles –} \\ \textbf{Cybersecurity engineering} \\ \textit{Microsoft STRIDE Threat Model}\end{tabular} & 
See \cite{stride} & See \cite{b42, stride} \\ \hline

\multirow{10}{*}{\begin{tabular}[c]{@{}l@{}}\textbf{ISA/IEC 62443 Security for}
\\
\textbf{Industrial Automation and Control} \\ \textbf{Systems}
\\
\textit{CIA Triad}\end{tabular}} & 
\textbf{Confidentiality} & Prevents unauthorized access to sensitive information. \\ \cline{2-3} 
& \textbf{Integrity} & Ensures the accuracy and consistency of data and system operations. \\ \cline{2-3} 
& \textbf{Availability} & Ensures systems function as intended without disruption over time. \\ \cline{2-3} 
& \textit{Access Control} & Restricts access to devices or information to authorized users only. \\ \cline{2-3} 
& \textit{Use Control} & Limits the usage of devices or data to authorized operations. \\ \cline{2-3} 
& \textit{Data Integrity} & Protects communication channels from unauthorized changes to data. \\ \cline{2-3} 
& \textit{Data Confidentiality} & Secures data from unauthorized access during transmission. \\ \cline{2-3} 
& \textit{Restrict Data Flow} & Controls data flow to prevent exposure to unauthorized entities. \\ \cline{2-3} 
& \textit{Timely Response to Event} & Ensures prompt responses to security incidents with appropriate actions. \\ \cline{2-3} 
& \textit{Resource Availability} & Guarantees resources remain accessible despite potential attacks. \\ \hline

\multirow{8}{*}{\begin{tabular}[c]{@{}l@{}}\textbf{IEEE 1547-3 Guide for Cybersecurity}\\ \textbf{of DERs Interconnected with} \\ \textbf{Electric Power Systems}\\ \textit{CIA Triad}\end{tabular}} & 
\textbf{Confidentiality} & Protects information from unauthorized disclosure. \\ \cline{2-3} 
& \textbf{Integrity} & Prevents and detects unauthorized data modification. \\ \cline{2-3} 
& \textbf{Availability} & Ensures systems remain operational and reliable. \\ \cline{2-3} 
& \textbf{Accountability} & Links actions to responsible entities for traceability. \\ \cline{2-3} 
& \textbf{Authentication} & Confirms the identity of communication participants. \\ \cline{2-3} 
& \textbf{Authorization} & Defines user permissions for data access and operations. \\ \cline{2-3} 
& \textbf{Non-repudiation} & Provides proof of origin for actions or commands. \\ \cline{2-3} 
& \textit{Cryptography} & Uses algorithms to secure data via encryption and hashing. \\ \hline
\multicolumn{3}{l}{\small $^{\text{\Cross}}$ Definitions have been paraphrased and condensed for readability. Exact definitions can be found in their respective standards.}
\end{tabular}
\end{table*}
\raggedbottom

Cybersecurity standards relevant to road vehicles, industrial automation and control systems, and power delivery electronics are also important considerations for EVSE software. 

\textit{Road vehicles.---}The standard for road vehicle cybersecurity, ISO/SAE 21434:2021, specifies requirements for threat modeling and recommends the Microsoft STRIDE Threat Model, as well as alternate frameworks: EVITA\footnote{https://doi.org/10.5281/zenodo.1188418}, TVRA\footnote{https://www.etsi.org/deliver/etsi\_ts/102100\_102199/10216501/05.02.03 \\ \_60/ts\_10216501v050203p.pdf}, and PASTA\footnote{https://threat-modeling.com/pasta-threat-modeling/}. 
We utilize STRIDE (as opposed to EVITA, TVRA, and PASTA) because STRIDE enables parsimony across Pillars II and III and simplifies the process of mapping threat scenarios to high-level security characteristics.

\textit{Industrial automation \& control systems.---}The series of industrial automation and control standards governing security, ISA/IEC 62443, aligns with the confidentiality, integrity, and availability (CIA) Triad\footnote{https://www.techtarget.com/whatis/definition/Confidentiality-integrity-and-availability-CIA}. The CIA triad are thus the characteristics we employ. The sub-characteristics from ISA/IEC 62443 are highlighted in Table III. 

\textit{Power delivery electronics.---}The cybersecurity standard relevant to power delivery electronics is the IEEE 1547-3 Guide for Cybersecurity of DERs Interconnected with Electric Power Systems. This standard combines multiple resources to define cybersecurity characteristics, including ISA/IEC 62443 \cite{b43}, the NIST Cybersecurity Framework \cite{b5}, the DHS US-CERT Cyber Security Evaluation Tool \cite{b78}, the MITRE ATT\&CK Framework, and the U.S. National Renewable Energy Laboratory DER Cybersecurity Framework \cite{b7}. 

Section 4.4.4.1 of IEEE 1547-3 lists many security requirements applicable to EVSE. We operationalize these requirements as HSQA characteristics (Fig. 1). This IEEE 1547-3 standard primarily utilizes the CIA Triad to map common cyberattacks to security characteristics, but includes additional important characteristics and a sub-characteristic outside of the CIA triad (Table III); the sub-characteristic of note is cryptography. The other sub-characteristics relevant to power delivery electronics are defined in ISA/IEC 62443 \cite{b63} (Table III).  

\section{Developing HSQA Models from Quality \& Security Characteristics}

Our current research operationalizes HSQA using the quality and security characteristics described in Section 4. Developing these HSQA models for EVSE requires extending our HSQA meta-model and specifying the characteristics as ``quality/security aspects" and the sub-characteristics as ``product factors" \cite{b53}. However, because HSQA is a tried and true technology with an extensible meta-model, the time from model concept to minimum viable product has been short. 

Operationalizing characteristics and sub-characteristics into an HSQA model is trivial. The challenge lies in ensuring the ``right" characteristics and sub-characteristics are selected for inclusion in the HSQA models for EVSE software.
As EVSE is relatively new, inherently complex, and a technology that operates at the nexus of multiple critical infrastructures, the characteristics and sub-characteristics identified above will almost certainly require refinement. We anticipate this refinement to be our greatest challenge. However, this challenge is one we have addressed in other domains. 

Our current work targets source and compiled code in EVSEs. Because the inputs to these HSQA models are the outputs of static analysis tools, an important consideration is that models can only aggregate results for measurable characteristics. Thus, it is imperative to find static analysis tools that have the content coverage for EVSE software; if such tools are not available, tooling is a threat to the internal and content validity of HSQA outputs. These threats to validity are concerns for EVSE HSQA models because tooling that is specific to EVSE infrastructure is of limited availability. In the absence of EVSE-specific tools, we are utilizing generic tools, including SonarQube\footnote{https://www.sonarsource.com/products/sonarqube/} for source code and CVE Binary Tool\footnote{https://github.com/intel/cve-bin-tool} for compiled code.

\section{Discussion}

HSQA models for EVSE software build upon the foundation laid by the Four-Interface \cite{b13} and CharIN EVSE threat models \cite{b14}. Our work incorporates these findings and provides the foundation for integrating them into deployable HSQA models. These HSQA models consider the identification, risk, and interdependencies of vulnerabilities to EVSE software and provide a means for evaluation throughout the software development lifecycle. HSQA thus advances the evaluation of software quality and security in EVSEs.

Our position acknowledges that EVSE software quality and security are not mutually exclusive. Rather, quality and security are interrelated, and are best addressed as such. Integrating software quality characteristics with EVSE cybersecurity characteristics enhances the overall security posture of EVSEs by directly measuring attack vectors to one of the most vulnerable digital attack surfaces in EVCI. 

Unlike traditional power grid and quality-of-service evaluations of EVSEs, which focus on external impacts and operational performance, HSQA delves into the integrity and robustness of the internal software components. HSQA measures code quality and security, thereby providing a holistic assessment of interdependent, complex cyber-physical software systems. 

In conclusion, we return to our position that HSQA is an excellent approach for the assessment of EVSE software. As EVSEs are a prime target and comprise a large and growing attack surface, cybersecurity solutions are needed. Here, our position includes a solution that is secure-by-design, tried-and-true, and comprehensive. HSQA will increase the quality and security posture of EVSE by offering a parsimonious solution for mitigating a wicked problem.

\section*{Acknowledgments}
This research was conducted with support from the U.S. Department of Homeland Security (DHS) Science and Technology Directorate (S\&T) and the Department of Energy (DOE) Pacific Northwest National Laboratory (PNNL) under contract IAA\# 70RSAT24KPM000022. Any opinions contained herein are those of the author and do not necessarily reflect those of DHS S\&T or DOE PNNL.

The PNNL AI Incubator Large Language Model was used in this document for spell-checking and grammatical enhancements.

\balance
\bibliographystyle{IEEEtran}
\bibliography{bibliography}

\end{document}